\documentclass[aps,prd,showpacs,superscriptaddress,nofootinbib,twocolumn]{revtex4}
\usepackage{graphicx}
\usepackage{amsmath}
\usepackage[sort&compress]{natbib}
\begin{document}
\title{The Diffuse Supernova Neutrino Background is detectable in Super-Kamiokande} 

\author{Shunsaku Horiuchi}
\affiliation{Department of Physics, School of Science, University of Tokyo, Tokyo 113-0033, Japan}
\affiliation{Center for Cosmology and Astro-Particle Physics, Ohio State University, Columbus, Ohio 43210}
\affiliation{Department of Physics, Ohio State University, Columbus, Ohio 43210}
\author{John F.~Beacom}
\affiliation{Center for Cosmology and Astro-Particle Physics, Ohio State University, Columbus, Ohio 43210}
\affiliation{Department of Physics, Ohio State University, Columbus, Ohio 43210}
\affiliation{Department of Astronomy, Ohio State University, Columbus, Ohio 43210}

\author{Eli Dwek}
\affiliation{Observational Cosmology Lab, NASA Goddard Space Flight Center, Greenbelt, MD 20771}

\date{\today}

\begin{abstract}
The Diffuse Supernova Neutrino Background (DSNB) provides an immediate opportunity to study the emission of MeV thermal neutrinos from core-collapse supernovae. The DSNB is a powerful probe of stellar and neutrino physics, provided that the core-collapse rate is large enough and that its uncertainty is small enough. To assess the important physics enabled by the DSNB, we start with the cosmic star formation history of Hopkins \& Beacom (2006) and confirm its normalization and evolution by cross-checks with the supernova rate, extragalactic background light, and stellar mass density. We find a sufficient core-collapse rate with small uncertainties that translate into a variation of $\pm 40$\% in the DSNB event spectrum. Considering thermal neutrino spectra with effective temperatures between 4--6 MeV, the predicted DSNB is within a factor \mbox{4--2} below the upper limit obtained by Super-Kamiokande in 2003. Furthermore, detection prospects would be dramatically improved with a gadolinium-enhanced Super-Kamiokande: the backgrounds would be significantly reduced, the fluxes and uncertainties converge at the lower threshold energy, and the predicted event rate is 1.2--5.6 events yr$^{-1}$ in the energy range 10--26 MeV. These results demonstrate the imminent detection of the DSNB by Super-Kamiokande and its exciting prospects for studying stellar and neutrino physics.   
\end{abstract}

\pacs{%
97.60.Bw, 
95.85.Ry, 
98.70.Vc  
}

\maketitle

\section{Introduction}

Core-collapse supernovae (CCSNe) occur at a rate of several per second in the Universe, each releasing a prolific $\sim$ $10^{58}$ neutrinos and antineutrinos. Their detection provides a rich bounty for stellar and neutrino studies, shown by the detection of the neutrinos from SN 1987A \cite{Hirata:1987hu,Bionta:1987qt,Hirata:1988ad,Bratton:1988ww,Arnett:1987iz,Bahcall:1987nx,Raffelt:1987yt,Barbieri:1988nh,Lattimer:1988mf,Arnett:1990au,Jegerlehner:1996kx,Lunardini:2004bj}. While a core-collapse supernova in the Milky Way would easily be detected in neutrinos, the occurrence rate is only $\lesssim$ 3 per century \cite{VanDenBergh:1991ke,Tammann:1994ev,Diehl:2006cf}. Proposed neutrino detectors should be able to detect supernovae up to 10 Mpc away with an occurrence rate of $\sim$ 1 per year \cite{Ando:2005ka,Kistler:2008us}. The vast majority of supernovae are therefore undetectable. However, the cumulative emission from all past core-collapse supernovae, which forms the Diffuse Supernova Neutrino Background (DSNB), has promising detection prospects \cite{BisnovatyiKogan:1984,Krauss:1983zn,Dar:1984aj,Woosley:1986aa,Totani:1995rg,Totani:1995dw,Malaney:1996ar,Hartmann:1997qe,Kaplinghat:1999xi,Fukugita:2002qw,Ando:2002zj,Ando:2002ky,Strigari:2005hu,Lunardini:2005jf,Yuksel:2007mn,Chakraborty:2008zp}. The Super-Kamiokande (SK) limit of $\phi(E_{\bar{\nu}_e} > 19.3 \,\mathrm{MeV}) < 1.2$ cm$^{-2}$ s$^{-1}$ \cite{Malek:2002ns} on the DSNB flux is already close to theoretical predictions \cite{Strigari:2005hu,Yuksel:2007mn}. 

Predicting the DSNB for a given supernova neutrino emission model requires knowledge of the rate of core-collapse supernovae. In the past, this was not well known, and various studies provided insights on the supernova rate \emph{from} the DSNB \cite{Totani:1995dw,Fukugita:2002qw,Ando:2004sb}. Indeed, the SK limit on the DSNB flux is strong enough to rule out some supernova rate evolution models, assuming a fiducial neutrino emission model. However, our understanding of the cosmic star formation history (CSFH) has been greatly augmented by improved direct measurements in different wavebands and redshifts (see, e.g., Refs.~\cite{Hopkins:2004ma,Hopkins:2006bw} and references therein). Cross-checks with other well-measured observables are now constraining, so that the CSFH is well determined by methods other than the DSNB. Thus it is both timely and important to study the prospects of the DSNB for probing stellar and neutrino physics. 

In this paper we start with the CSFH of Hopkins \& Beacom (hereafter HB06 \cite{Hopkins:2006bw}) which is based on a compilation of recent data, and assess it by cross-checks, with the aim of evaluating the uncertainties that are carried forward into DSNB predictions through the core-collapse supernova rate. We henceforth refer to these as the \underline{\it astrophysical} inputs and uncertainties on the DSNB. On the other hand, we refer to the supernova neutrino emission and neutrino properties as the \underline{\it emission} inputs and uncertainties on the DSNB.

As cross-check material we consider measurements of the rate of core-collapse supernovae, which have been significantly updated \cite{Cappellaro:2004ti,Dahlen:2004km,Botticella:2007er,Smartt:2008zd,Dahlen:2008aa}, the extragalactic background light, which records the total stellar emission over all time (for a recent review, see, e.g., Ref.~\cite{Hauser:2001xs}), and, finally, the stellar mass density (see e.g., Ref.~\cite{Wilkins:2008be} and references therein). While our approach is similar to previous studies such as Ref.~\cite{Strigari:2005hu}, we perform novel checks and with higher precision. In particular, our analysis delivers fiducial inputs with unprecedentedly small uncertainties.

Using our constrained astrophysical inputs, we find that the DSNB uncertainty is dominated by the emission inputs, demonstrating the potential to study stellar and neutrino physics using the DSNB. Furthermore, taking into account both astrophysical and emission inputs, we find that the predicted DSNB is within at most a factor \mbox{$\sim$ 4} of the SK limit set in 2003. For a 6 MeV thermal spectrum, typical of scenarios with neutrino mixing, this factor reduces to $\sim$ 2. At the lower detection threshold energy for a gadolinium-enhanced SK \mbox{($\sim 10$ MeV} \cite{Beacom:2003nk,Yuksel:2005ae}), the combined uncertainty on the predicted DSNB is remarkably only a factor 2, with a event rate of 1.2--5.6 events yr$^{-1}$ in the energy range 10--26 MeV. Thus, the core-collapse rate and the neutrino emission per supernova are large enough to allow the imminent detection of the DSNB in SK. The successful detection will confirm the ubiquitous emission of neutrinos from core-collapse supernovae and initiate the much-anticipated study of stellar and neutrino physics, while a non-detection would require new stellar or neutrino physics. 

The paper is organized as follows. In Sections \ref{sec:astrophysics} and \ref{sec:dsnb} we discuss the astrophysical and emission inputs for the DSNB. We discuss the DSNB detectability and make future predictions for a gadolinium-enhanced SK in Section \ref{sec:sk}, and finish with conclusions in Section \ref{sec:conclusion}. We adopt the standard $\Lambda$CDM cosmology with $\Omega_m=0.3$, $\Omega_\Lambda=0.7$, and $H_0=70$ km s$^{-1}$ Mpc$^{-1}$.

\section{Astrophysical inputs \label{sec:astrophysics}}
For a given supernova neutrino spectrum, the key input for calculating the DSNB flux is the history of the rate of CCSNe (this includes Type II and the subdominant Type Ib/c supernovae). The CCSN rate is directly related to the birth and death rate of massive stars. In recent years, data on the star formation rate (SFR) have improved both in breadth and sophistication, leading to an unprecedented understanding of the CSFH. In this section, we start with the latest CSFH and cross-check it with the measured CCSN rate, extragalactic background light (EBL), and stellar mass. Importantly, these checks probe different stellar life phases and are sensitive to different stellar masses, making them complementary. Thus we are able to self-consistently assess the astrophysical inputs for the DSNB.

\subsection{Cosmic star formation history \label{sec:sfr}}
The SFR is most often derived from measurements of living massive stars. The measured luminosities, together with knowledge of their masses and lifetimes, gives their birth rates. Since the most massive stars have the shortest lifetimes, they provide a measure of the most recent star formation activity. In practice, the observed luminosities are corrected for dust and the total SFR (over the entire stellar mass range) is derived by extrapolation to lower masses using the initial mass function (IMF). The calibration is done by use of a stellar population code that calculates the radiative output from a population of stars given an IMF. 

We compute calibration factors using the PEGASE.2 stellar population code \cite{Fioc:1997sr}, which contains a careful treatment of stellar physics. We assume constant SFR bursts of 10$^8$ years, a close binary fraction of 0.05, evolutionary tracks with stellar winds, the supernova model B of Woosley \& Weaver \cite{Woosley:1995ip}, and a constant metallicity of $Z=0.02$ (i.e., solar). The parameter that the results are the most sensitive to is the star formation duration. For integrated measurements of galaxies, it is usually appropriate to assume the SFR has remained constant over time scales that are long compared to the lifetimes of the dominant UV emitting population ($\sim 10^8$ years) \cite{Kennicutt:1998zb}; however, there are calibration uncertainties of a few tens of percent (see Table \ref{table:calibration}). In contrast, studies suggest that the mean metallicity of star formation gas is close to the solar value for redshifts of a few \cite{Panter:2002ed,Panter:2008vw}, resulting in calibration uncertainties of ten percent or less. The other parameters yield variations of only a few percent. We note the agreement of PEGASE.2 outputs with other stellar population synthesis codes demonstrated in e.g., Refs.~\cite{Bruzual:2003tq,Fardal:2006sd}. 

\begin{figure}[t]
\includegraphics[width=3.25in]{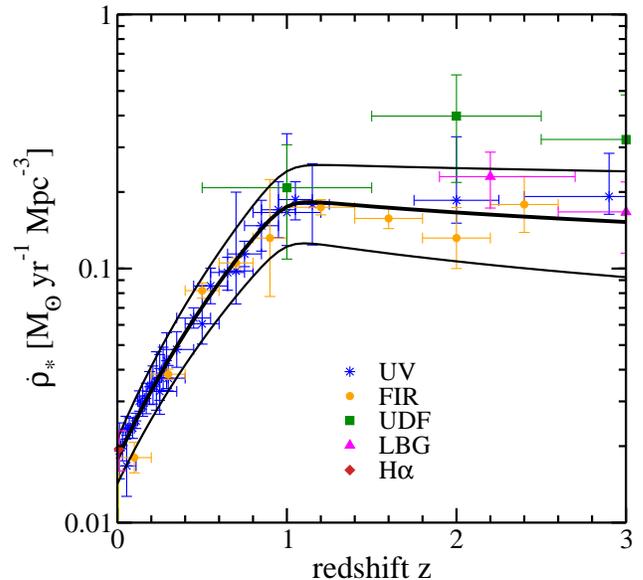}
\caption{\label{fig:sfr} Redshift evolution of the comoving SFR density. Data derived from various indicators are included as labeled, all scaled to a Salpeter IMF \cite{Salpeter:1955it}. The majority are from the compilation of HB06 \cite{Hopkins:2006bw}, with additional data from LBG \cite{Reddy:2008rj} and H$\alpha$ measurements \cite{James:2008pa}. We plot the fiducial CSFH (thick solid curve) and our generous adopted uncertainty range (thin solid curves). The curves take into account SFR data at higher redshifts \cite{Hopkins:2006bw,Yuksel:2008cu} that are not shown here, where we focus on the lower redshifts most relevant for the DSNB.}
\end{figure}

\begin{table}
\caption{Calibration factors, $f_{\rm bol}=L_{\rm bol}/\dot{\rho}_*$ and $f_{\rm UV}=L_{\rm UV}/\dot{\rho}_*$, calculated using PEGASE.2 \cite{Fioc:1997sr}, for three IMFs \cite{Salpeter:1955it,Kroupa:2000iv,Baldry:2003xi} and three epochs (yr). For a particular IMF, the star formation duration causes an uncertainty of a few tens of percent about the central $10^8$ yr value. This contributes to the scatter in the SFR for a given IMF. Note $f_{\rm UV}$ is determined at $\lambda = 0.2$ $\mathrm{\mu m}$ since the UV output is nearly constant over wavelength for the durations considered.}\label{table:calibration}
\begin{ruledtabular}
\begin{tabular}{lcccccccccc} 
& \multicolumn{2}{c}{IMF slope\footnotemark[1]} && \multicolumn{3}{c}{$f_\mathrm{bol}$\footnotemark[2]} && 
\multicolumn{3}{c}{$f_\mathrm{UV}$\footnotemark[3]} \\
IMF & $\xi_1$ & $\xi_2$ && $10^7$ & $10^8$ & $10^9$ && $10^7$ & $10^8$ & $10^9$ \\
\hline 
Salpeter (1955) & 2.35 & 2.35 && 4.3 &  6.5 &  8.6 && 5.1 &  7.8 &  8.9 \\
Kroupa (2001)   & 1.3  & 2.3  && 6.7 &  9.9 & 13   && 7.9 & 12   & 13   \\
BG (2003)       & 1.5  & 2.15 && 8.5 & 12   & 15   && 9.9 & 14   & 16   \\
\end{tabular} 
\footnotetext[1]{$\psi(M) \propto M^{-\xi_1}$ for 0.1--0.5 $\mathrm{M_\odot}$ and $M^{-\xi_2}$ for 0.5--100 $\mathrm{M_\odot}$}
\footnotetext[2]{in $10^9$ $\mathrm{L_\odot \, (M_\odot/yr)^{-1}}$}
\footnotetext[3]{in $10^{27}$ $\mathrm{ erg \, s^{-1} \, Hz^{-1} (M_\odot/yr)^{-1}}$}
\end{ruledtabular}
\end{table} 

We calculate results for three IMFs: the traditional steeper Salpeter IMF \cite{Salpeter:1955it}, an intermediate Kroupa IMF \cite{Kroupa:2000iv}, and a shallower Baldry-Glazebrook (BG) IMF \cite{Baldry:2003xi}. We define the IMF as $\psi(M)=\mathrm{d}N/\mathrm{d}M$ so that $\psi(M) \mathrm{d}M$ gives the number of stars in the mass range $M$ to $M+\mathrm{d}M$. The slopes $\psi(M) \propto M^{-\xi}$ are shown in Table \ref{table:calibration}. 
 
We take the SFR compilation and fit of HB06 as our starting point and add recent data. The HB06 data consist of various SFR indicators, including UV measurements from SDSS \cite{Baldry:2005df}, GALEX \cite{Schiminovich:2004km,Arnouts:2005aa} and COMBO17 \cite{Wolf:2002ks}, far-infrared (FIR) from Spitzer \cite{PerezGonzalez:2005bt}, and high redshift measurements from the Hubble Ultra Deep Field (UDF) \cite{Thompson:2006bb}; to this we add recently-derived data from Lyman Break Galaxies (LBG) \cite{Reddy:2008rj} and from H$\alpha$ emission \cite{James:2008pa}. These are shown in Fig.~\ref{fig:sfr} as a function of redshift, all scaled to a Salpeter IMF. Assuming the Kroupa or BG IMF results in values that are lower by an overall factor $\simeq$ 0.66 and $\simeq$ 0.55, respectively. 

Fig.~\ref{fig:sfr} demonstrates the overall consistency between SFR densities estimated by different indicators, over a wide range of redshifts. In general, the systematic scatter due to different indicators dominates over the formal uncertainties from the calibrations of each indicator. However, at low redshifts ($z < 1$), the scatter decreases and approaches the calibration uncertainties. To further constrain the CSFH in the future would require calibration uncertainties to be examined. We note that redshift dependent dust corrections have been applied. At $z<1$, the UV and FIR measurements are combined, while for $1<z<3$, a constant dust correction is made; see Ref.~\cite{Hopkins:2004ma} for further details.

To compare the CSFH to other observables, it is useful to define an analytic fit. We adopt a continuous broken power-law as in Ref.~\cite{Yuksel:2008cu},
\begin{equation}\label{fit}
\dot{\rho}_*(z) = \dot{\rho}_0 \left[ 
(1+z)^{\alpha \eta} + 
\left( \frac{ 1+z }{B} \right)^{\beta \eta} + 
\left( \frac{ 1+z }{C} \right)^{\gamma \eta} \, 
\right]^{1/\eta},
\end{equation} 
where $\dot{\rho}_0$ is the normalization, $B$ and $C$ encode the redshift breaks, the transitions are smoothed by the choice $\eta \simeq -10$, and $\alpha$, $\beta$ and $\gamma$ are the logarithmic slopes of the low, intermediate, and high redshift regimes, respectively. The constants $B$ and $C$ are defined as
\begin{eqnarray}
B &=& (1 + z_1)^{1-\alpha/\beta}, \\
C &=& (1 + z_1)^{(\beta-\alpha)/\gamma} (1 + z_2)^{1-\beta/\gamma},
\end{eqnarray}
where $z_1$ and $z_2$ are the redshift breaks. We adopt the fiducial CSFH fit obtained by combining the HB06 compilation with new measurements at high redshift derived from gamma-ray bursts \cite{Kistler:2007ud,Yuksel:2008cu} (note that high-redshift data are not shown in Fig.~\ref{fig:sfr}). We define a generous envelope that takes into account the scatter in the data, as shown in Fig.~\ref{fig:sfr}. The parameters are given in Table \ref{table:fit}. 

\begin{table} 
\caption{CSFH parametric fit to the form of Eq.~(\ref{fit})}\label{table:fit}
\begin{ruledtabular}
\begin{tabular}{ccccccc}
Analytic fits\footnotemark[1] & $\dot{\rho}_0$ & $\alpha$ & $\beta$ & $\gamma$ & $z_1$ & $z_2$ \\
\hline 
Upper    & 0.0213 & 3.6 & -0.1 & -2.5 & 1 & 4 \\
Fiducial & 0.0178 & 3.4 & -0.3 & -3.5 & 1 & 4 \\
Lower    & 0.0142 & 3.2 & -0.5 & -4.5 & 1 & 4 \\
\end{tabular}
\end{ruledtabular}
\footnotetext[1]{Shown for the Salpeter IMF. For the Kroupa and BG IMFs the normalization $\dot{\rho}_0$ decreases by a factor $\simeq$ 0.66 and $\simeq$ 0.55, respectively; the overall shape is not greatly affected (see, e.g., Table 2 of Ref.~\cite{Hopkins:2006bw}). Units of $\rho_0$ are in $\mathrm{M_\odot \, yr^{-1} \, Mpc^{-3}}$.}
\end{table} 

\subsection{Rate of core-collapse supernovae \label{sec:CCSN}}
The number of stars per unit mass undergoing core collapse is dependent on the IMF and mass range of stars that lead to core collapse. The predicted comoving CCSN rate history, $R_{\rm CCSN}(z)$, follows directly from the CSFH,
\begin{equation}\label{CCSNrate}
R_{\rm CCSN}(z) = \dot{\rho}_*(z)
\frac{\int_{8}^{50}\psi(M)dM}{\int_{0.1}^{100} M \psi(M)dM},
\end{equation}
where the ratio of integrals is 0.0070/$\mathrm{M_\odot}$, 0.0109/$\mathrm{M_\odot}$, and 0.0132/$\mathrm{M_\odot}$ for the Salpeter, Kroupa, and BG IMFs, respectively.

The lower mass threshold is the most important parameter for calculating $R_{\rm CCSN}$. In general, it is difficult to predict accurately from theory, because stellar properties change rapidly between \mbox{$\sim$ 6--10 $M_\odot$}; it is usually assumed to be 8 $\mathrm{M_\odot}$. On the other hand, the mass can be determined from direct identifications of progenitor stars from pre-explosion imaging. The most recent study places the mass threshold for Type II-P supernovae at $8.5^{+1}_{-1.5}\,\mathrm{M_\odot}$ \cite{Smartt:2008zd}. The uncertainties would translate to an uncertainty of about ten percent in $R_{\rm CCSN}$. However, one should keep in mind that explosions of O-Ne-Mg cores may result from a lower mass range that leads to a different variety of CCSN. The upper mass limit is less important because of the steep IMF, and only affects the results at the percent level. Theoretically, prompt black hole formation occurs above some critical mass; the precise value depends strongly on properties such as rotation and metallicity, but is thought to lie around 30--50 $\mathrm{M_\odot}$ \cite{Fryer:1999mi}. Such explosions will generally not produce optical supernovae but will likely produce neutrinos. Observationally, a progenitor mass of at least 40 $\mathrm{M_\odot}$ has been claimed to have been inferred from a neutron star in a very young cluster \cite{Muno:2005xn}. 

\begin{figure}[t]
\includegraphics[width=3.25in]{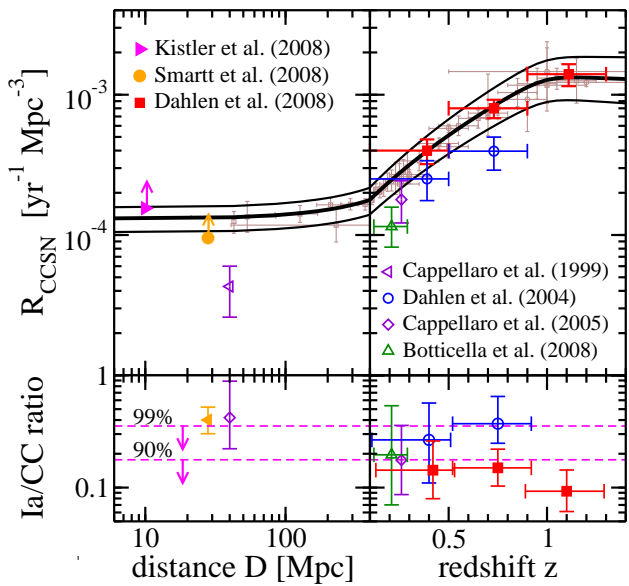}
\caption{\label{fig:CCSNrate} Evolution of the core-collapse supernova rate (top panel) and Type Ia to CCSN ratio (bottom panel), shown against distance in Mpc (for $z<0.1$) and redshift (for $z>0.1$). Data as labeled; filled symbols indicate data or limits we adopt, and empty points indicate published rates we treat as lower limits. Error bars show statistical errors, except for horizontal bars on Dahlen et al.~points that show bin size. In the top panel, predictions from the CSFH are shown by the solid curve, where the uncertainty band propagates from Fig.~\ref{fig:sfr}. The SFR measurements, scaled to a CCSN rate, are shown in the background (small light brown squares). The BG IMF has been adopted; using other IMFs causes differences of only a few percent. In the bottom panel, the dashed lines denote conservative upper limits from the non-observation of any Type Ia supernovae within 10 Mpc \cite{Kistler:2008us}. The 10 Mpc point has been very conservatively selected from Ref.~\cite{Kistler:2008us}, and the rate has been further corrected downwards (see text).}
\end{figure}

It should be noted that the predicted $R_{\rm CCSN}$ is largely insensitive to the IMF. This is because the effects of the IMF nearly cancel out between $\dot{\rho}_*$ and the integrals in Eq.~(\ref{CCSNrate}). For instance, a shallower IMF contains relatively more massive stars but has a smaller $\dot{\rho}_*$. Numerically, e.g., $R_{\rm CCSN}^{Sal} \approx 0.96 R_{\rm CCSN}^{BG}$. This is a natural outcome because both the CCSN and the SFR indicators are associated with the same massive stars. Thus, the CCSN rate allows an almost IMF-independent check of the CSFH.

In the top panel of Fig.~\ref{fig:CCSNrate} we show the predicted CCSN rate, using the BG IMF, for the fiducial CSFH (thick solid curve) as well as the upper and lower CSFH (thin solid curves). Using the Kroupa or Salpeter IMF produces variations of only a few percent, as expected. For data, we plot rates derived from supernova compilations \cite{Kistler:2008us,Smartt:2008zd}, extended supernova surveys using the Hubble Space Telescope \cite{Dahlen:2004km,Dahlen:2008aa}, searches in nearby galaxies \cite{Cappellaro:1999qy,Cappellaro:2004ti}, and the Southern inTermediate Redshift ESO Supernova Search (STRESS) \cite{Botticella:2007er}. We also show in the background the SFR measurements scaled via Eq.~(\ref{CCSNrate}) to a CCSN rate using the BG IMF. 

For the local Universe, CCSN rates are derived from compilations of supernovae occurring within 10 Mpc over a time period of 10 years \cite{Kistler:2008us}, and within 28 Mpc over a time period of 10.5 years \cite{Smartt:2008zd}. These compilations include Types Ia, II, Ib, and Ic supernovae, the latter three of which are used to derive the CCSN rate. Due to the known incompleteness of the compilations, both should be treated as lower limits on the CCSN rate. In addition, for the 10 Mpc point, we conservatively exclude sources that might inadvertendly be outside 10 Mpc and we further exclude transients of peculiar origin (e.g., SN 2002kg, SN 2008S and the 2008 transient in NGC 300) \cite{Thompson:2008sv}, which reduces the number of CCSN to 13. Also, the 10 Mpc point has been corrected downwards for the local increase in SFR density. We apply a simple correction assuming the SFR density traces the galaxy density; the galaxy over-density in 10 Mpc is about a factor 2 according to galaxy catalogues \cite{Karachentsev:2004aa} (see also Fig.~2 of Ref.~\cite{Blanton:2000dr}). The correction does not affect the Type Ia to CCSN ratio that we discuss below.

Since the lifetimes of massive stars are short, the shape of the CCSN rate evolution must follow that of the CSFH. On the other hand, the normalization needs to be checked. We find excellent agreement between predictions from the CSFH and the most recent data from the survey of Dahlen et al.~using the Hubble Space Telescope \cite{Dahlen:2008aa}. These data have been derived by periodically observing the same patch of sky, locating supernovae within a volume only limited by flux. The new data (red solid squares) \cite{Dahlen:2008aa} are updates of their previous data (blue empty circles) \cite{Dahlen:2004km}, and have much better statistics and a more detailed analysis (we plot the data which have been corrected for dust). We find further agreement with recently published supernova compilations within 10 Mpc \cite{Kistler:2008us} and 28 Mpc \cite{Smartt:2008zd}, even though they are certainly incomplete, especially in the Southern sky, confirming the minimal normalization over a wide distance scale. 

On the other hand, data from Cappellaro et al.~\cite{Cappellaro:1999qy,Cappellaro:2004ti} and STRESS \cite{Botticella:2007er} fall short of the trend of other data. Unlike in Dahlen et al., these surveys search for supernovae by periodically observing a pre-selected sample of galaxies in a given field. While a large number of galaxies are selected, even this is likely incomplete, as small galaxies are often under-sampled. In addition, as the authors clarify, host galaxy extinction is an important uncertainty on supernova surveys \cite{Botticella:2007er}. The surveys of Cappellaro et al.~do not include host galaxy extinction, and as the authors state, they most likely have too few CCSN \cite{Cappellaro:2004ti}. 

The ratio of Type Ia to CCSN also assists in assessing the results of supernova surveys. Due to their brighter nature, Type Ia supernovae are more easily detected, so that the ratio of Type Ia to CCSN will increase if CCSN are increasingly missed. In the bottom panel of Fig.~\ref{fig:CCSNrate}, we show the ratio of raw supernova counts for Ref.~\cite{Smartt:2008zd}, and the ratio of reported Type Ia and CCSN rates for other studies. From the non-observation of any Type Ia supernovae within 10 Mpc -- of the supernovae considered in the 10 Mpc compilation, none were Type Ia \cite{Kistler:2008us} -- we determine 90\% and 99\% upper limits on the ratio, shown by dashed lines. We apply these upper limits over all distances and redshifts; although the ratio in principle evolves with time due to Type Ia time-delay effects, the evolution is not strong for global samples and $z \lesssim 1$ (see, e.g., Refs.~\cite{Neill:2007kg,Pritchet:2008np}), and moreover, a delayed component to the Ia rate relative to the CCSN rate would only make the limits stronger by increasing the local ratio. We therefore use the dashed lines as a conservative indicator for the true ratio, for simplicity. We see from Fig.~\ref{fig:CCSNrate} that more CCSN are being missed in the data we treat as lower limits (empty symbols). For example, the large ratio of Ref.~\cite{Cappellaro:1999qy} predicts $\sim 5$ Type Ia within 10 Mpc, in disagreement with observations \cite{Kistler:2008us}. Alternatively, taking the 90\% upper limit as described above and respecting the reported Type Ia supernova rates, the CCSN rates (empty symbols) increase to values more in line with theoretical predictions from the CSFH. 

We therefore conclude the data confirm the CSFH normalization over a large range of distances. Moreover, the range is precisely that of interest for the DSNB (almost all of the detectable flux comes from $z < 1$).

As an aside, while predictions include all core collapses, including those with little or no optical signals due to the prompt formation of a black hole or due to dust obscuration, the data are derived from optical supernovae only. The excellent agreement between prediction and data suggests that dark core collapses comprise a minority of all core collapses. Alternatively, the fraction of optically-dark but neutrino-bright collapses can be increased by adjusting the CSFH and mass range for core collapse; however, this would be accompanied by a correspondingly larger DSNB flux, which is constrained by the tight experimental limits discussed below. 

We note in this context the red supergiant problem noted by Smartt et al.~\cite{Smartt:2008zd}. They find a shortage of Type II-P supernovae associated with red supergiants in the mass range 17--25 $\mathrm{M_\odot}$, despite clear evidence that these massive supergiants exist. These stars constitute $\sim$ 10\% of all massive stars leading to core collapse, and although slightly smaller in mass than theoretical predictions, they could be dark core collapses. Monitoring nearby red supergiants can reveal the occurrence rate of dark core collapses, as proposed by \cite{Kochanek:2008mp}.

\subsection{Extragalactic background light}
It is widely accepted that the EBL is the record of the total stellar emission over all time. The observed EBL is dominated by two peaks of comparable energy density. The first peak in the optical to near infrared (NIR) is powered by direct starlight, while the second in the FIR is dominated by starlight that is absorbed and re-emitted by dust. Since circumstellar dust absorbs most efficiently in the UV, the power source of the FIR peak is primarily massive stars. The relative energy densities of the two peaks is a testament of the heavy dustiness of the present and past Universe (for a review, see e.g., Ref.~\cite{Hauser:2001xs}).

We check the CSFH by using the \emph{total observed EBL}. In principle, if one had precise knowledge of dust -- its quantity, properties and evolution -- one could use the spectral shape of the EBL to probe different stellar mass regimes. However, a full treatment is beyond the scope of this paper and we simply exploit the total EBL as a calorimeter of the total energy radiated by stars. Although this includes contributions from less massive stars, half of the total EBL is powered by stars with masses $M \gtrsim 3 \, \mathrm{M_\odot}$ for the BG IMF. 

\begin{figure}
\includegraphics[width=3.25in]{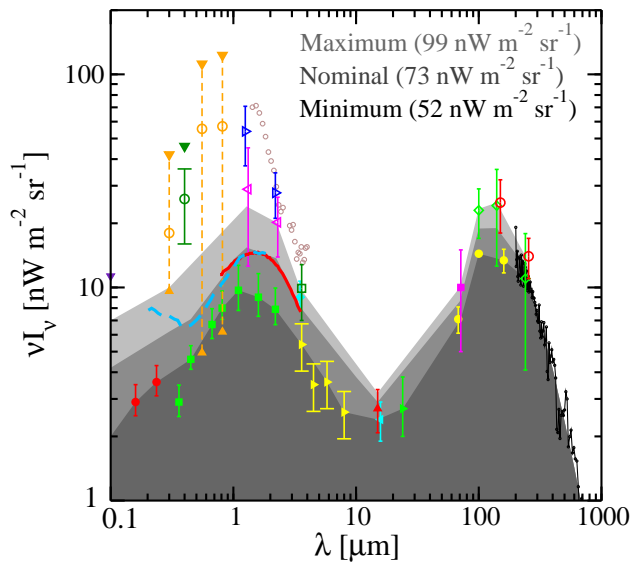}
\caption{\label{fig:ebl}The observed EBL spectrum. Filled symbols are based on integrated galaxy counts, while empty symbols and the black lines at high wavelength represent absolute measurements. In the UV to NIR, we also show the gamma-ray upper limits placed by H.E.S.S. (red solid curve) \cite{Aharonian:2005gh}, MAGIC (blue dashed curve) \cite{Aliu:2008ay}, and Edelstein et al.~2000 (filled downward triangle at 0.1 $\mu$m) \cite{Edelstein:2000aa}. We show three shaded regions representing the minimum (dark shading), nominal (light shading) and maximum (lightest shading) EBL. The integrated total EBL from these \medskip  are listed at the top-right. \\[-2mm]
\indent The integrated galaxy counts shown include (in order of increasing $\lambda$) Gardner et al.~2000 (red cicle) \cite{Gardner:2000aa}, Madau \& Pozzetti 2000 (green square) \cite{Madau:1999yh}, Levenson \& Wright 2008 (blue diamond) \cite{Levenson:2008aa}, Fazio et al.~2004 (yellow right-pointed triangle) \cite{Fazio:2004aa}, Metcalfe et al.~2003 (red up-pointed triangle) \cite{Metcalfe:2003zi}, Elbaz et al.~2002 (blue down-pointed triangle) \cite{Elbaz:2002vd}, Papovich et al.~2004 (green right-pointed triangle) \cite{Papovich:2004aa}, Dole et al.~2006 (yellow circle) \cite{Dole:2006aa}, and Frayer et al.~2006 (purple square) \cite{Frayer:2006aa}. Absolute measurements shown include (in order of increasing $\lambda$) Cambresy et al.~2001 (dark blue right-pointed triangle) \cite{Cambresy:2001aa}, Wright 2001 (purple left-pointed triangle) \cite{Wright:2000aa}, Dwek \& Arendt 1998 (dark green square) \cite{Dwek:1998aa}, Lagache et al.~2000 (green diamond) \cite{Lagache:2000aa}, and Hauser et al.~1998 (red cirlce) \cite{Hauser:1998aa}. For the results of Bernstein et al.~\cite{Bernstein:2002aa,Bernstein:2002ab,Bernstein:2005aa,Bernstein:2007aa}, we show $2\sigma$ upper limits (filled downward-pointing triangles), nominal values (empty circles), and lower limits (filled upward-pointing triangles), all connected by dashed lines. We show the diffuse sky measurements of Matsumoto et al.~2005 (small empty circles) \cite{Matsumoto:2005aa}, and the FIR measurements by Fixsen et al.~1998 (filled black circles) \cite{Fixsen:1998kq} connected by black lines.}
\end{figure}

Direct measurements of the EBL in the optical to NIR are complicated by foreground contamination \cite{Hauser:2001xs}. Indirect limits have been placed from integrated galaxy counts and from the opacity of the Universe to gamma rays using distant TeV blazars. We caution however that gamma-ray constraints depend on the assumed intrinsic blazar spectral index, so the result may be weakened in light of unknown acceleration mechanisms \cite{Dwek:2004pp,Stecker:2007aa,DeAngelis:2008sk,Razzaque:2008te,Stecker:2008aa}. In Fig.~\ref{fig:ebl} we present the latest data on the EBL, including direct measurements (empty symbols) \cite{Bernstein:2002aa,Bernstein:2002ab,Bernstein:2005aa,Bernstein:2007aa,Cambresy:2001aa,Wright:2000aa,Dwek:1998aa,Lagache:2000aa,Hauser:1998aa}, lower limits from galaxy counts (filled symbols) \cite{Gardner:2000aa,Madau:1999yh,Levenson:2008aa,Fazio:2004aa,Metcalfe:2003zi,Elbaz:2002vd,Papovich:2004aa,Dole:2006aa,Frayer:2006aa}, and upper limits from gamma-ray attenuation (thick curves) \cite{Aharonian:2005gh,Aharonian:2007aa,Aliu:2008ay}. 

The systematic uncertainties on galaxy photometry and zodiacal light subtraction suggests that a curve between the measurements and counts represents the most appropriate assumption for the total observed EBL. We show three EBL regions: the minimum, which essentially traces the galaxy counts, the nominal, which respects the gamma-ray constraints, and the maximum, which lies within most of the data error bars. We do not consider the 1--4 micron diffuse sky emission detected by IRTS \cite{Matsumoto:2005aa} to be of extragalactic origin, since it is most likely caused by emission from interplanetary dust particles \cite{Dwek:2005dj}. The total EBL from these regions are 52, 73 and 99, all in units of $\mathrm{nW \, m^{-2} \, sr^{-1}}$. Our estimates are slightly higher than those of Refs.~\cite{Madau:1999yh,Madau:2000yn}, who find a best estimate of 60 $\mathrm{nW \, m^{-2} \, sr^{-1}}$. On the other hand, they are more in line with recent estimates of 60--93 $\mathrm{nW \, m^{-2} \, sr^{-1}}$ \cite{Gispert:2000np}, 45-170 $\mathrm{nW \, m^{-2} \, sr^{-1}}$ \cite{Hauser:2001xs}, and 50-129 $\mathrm{nW \, m^{-2} \, sr^{-1}}$ \cite{Fardal:2006sd}.

To calculate the EBL for a given CSFH, we need the spectral luminosity density $\epsilon(\nu,z)$ as a function of $z$, measured since the epoch $z_*$ when stars first turned on. Assuming the EBL arises dominantly from stellar radiation, $\epsilon$ is given by \cite{Dwek:1998ab}
\begin{equation}
\epsilon(\nu,z) = \int^{t_z}_{t_*} \dot{\rho}_*(t) \mathrm{d}t
\int^{M(t^\prime)}_{0.1} L(\nu,M,t^\prime) \psi(M) \mathrm{d}M, 
\end{equation}
where $L(\nu,M,t^\prime)$ is the luminosity per unit mass of a star of initial main-sequence mass $M$ at time $t^\prime = t_z - t$, $\psi(M)$ is the IMF in the range 0.1--100 $\mathrm{M_\odot}$ and $M(t^\prime)$ is the initial main-sequence mass of a star with a lifetime of $t^\prime$. We calculate this quantity using the PEGASE.2 stellar population code, with the same assumptions as described in Section \ref{sec:sfr}, but with evolution of metallicity, with an initial value of $Z=0.001$ at $z=10$. The specific EBL intensity $I(\nu)$ at the observed frequency $\nu$ is then the integral of $\epsilon(\nu^\prime,z)$, from the comoving volume element at $z$, over redshifts,
\begin{equation}
I(\nu)= \frac{c}{4 \pi} \int^{z_*}_0
\epsilon(\nu^\prime,z) \left| \frac{\mathrm{d}t}{\mathrm{d}z} \right| \mathrm{d}z,
\end{equation}
where $\nu^\prime=\nu(1+z)$ is the frequency at emission and $|\mathrm{d}z/\mathrm{d}t|=H_0(1+z) [\Omega_m(1+z)^3+\Omega_\Lambda]^{1/2}$. 

The values of the calculated total EBL are $95^{+39}_{-30}$ $\mathrm{nW \, m^{-2} \, sr^{-1}}$, $88^{+36}_{-28}$ $\mathrm{nW \, m^{-2} \, sr^{-1}}$, and $78^{+31}_{-24}$ $\mathrm{nW \, m^{-2} \, sr^{-1}}$ for the Salpeter, Kroupa, and BG IMFs, respectively. While the CSFH differs by almost a factor 2 between the Salpeter and BG IMFs, the calculated total EBL differs by much less, because the total EBL is dominated by relatively higher mass stars. On the other hand, steeper IMFs have more low mass stars which live long and pile up, which works to increase the total EBL. These dependencies on the IMF have in fact been studied by various authors to constrain the IMF (e.g., Ref.~\cite{Fardal:2006sd}). Of importance to us is the agreement between observations and predictions, when calculated using recent IMFs with shallower slopes and suppression at the lower mass end (BG IMF). 

We note that this result contrasts with a recent study which found predictions that were smaller than observations \cite{Fardal:2006sd}. In their study, the predicted EBL was $\sim 50$ $\mathrm{nW \, m^{-2} \, sr^{-1}}$ for the BG IMF and they estimated the observed EBL to be $77$ $\mathrm{nW \, m^{-2} \, sr^{-1}}$. Our study differs in three aspects. First, by including the gamma-ray constraints, the updated observed EBL is slightly smaller. Second, the updated CSFH is somewhat larger. Finally, we include consistent evolution of metallicity in our calculations, whereas the authors in \cite{Fardal:2006sd} assumed a constant $Z=0.02$ over all redshifts. Lower metallicity leads to less mass loss and hence higher time-integrated radiative output. These three factors result in us obtaining better agreement.

\subsection{Total stellar mass \label{sec:mass}}
Integration of the CSFH over redshift with appropriate corrections for stellar mass loss yields the stellar mass density. This quantity can be independently measured using galaxy surveys, which are often coupled to NIR observations as a proxy for stellar mass. Therefore, it also provides an independent check of the CSFH. The comparison has the property of probing a lower stellar mass range than the EBL. Numerous studies have been made, with results varying from good agreement \cite{Cole:2000ea,Fontana:2004aa,Arnouts:2007wi} to the CSFH over-producing stars \cite{Cole:2005sx,Hopkins:2006bw}. In these comparisons, the IMF plays a critical role \cite{Kroupa:2007wz}. A recent detailed study shows that the CSFH and observations of stellar mass density are in good agreement for redshifts $z \lesssim 0.7$ \cite{Wilkins:2008be}. In a subsequent paper, the authors find that if the IMF is constant in time, the best-fit IMF slope is 2.15 \cite{Wilkins:2008sa}, which is the same as our adopted BG IMF. Although the check becomes complicated by large scatter in measurements at redshifts above 0.7, the studies illustrate the overall consistency of the CSFH and stellar mass in the redshift of our interest and the preference for a shallow IMF.

\subsection{Consistency of the CSFH and observations}
Besides systematic uncertainties that contribute to the scatter in the CSFH, another issue is the CSFH normalization uncertainty arising from dust correction. As we discuss below, the true CSFH cannot be smaller by a factor at most $\sim$ 2, and even this seems unlikely.

Various studies have shown that the EBL is dominated by stars, with little contribution from non-nucleosynthesis energy sources such as active galactic nuclei \cite{Hopkins:2005fb,Fardal:2006sd}. Hence, we infer the minimum CSFH from the minimum observed EBL, $\approx$ $50 \, \mathrm{nW \, m^{-2} \, sr^{-1}}$. Requiring the minimum observed EBL to be explained, the true CSFH could be at most a factor $95/50 \sim 2$ smaller than the fiducial CSFH. 

However, this would require several unlikely changes. First, the Salpeter IMF is disfavored by other observables such as the stellar mass density. Second, the dust correction applied to the CSFH is typically a factor 2--3. Invoking a true dust-correction of 1--1.5 would imply almost negligible true dust, in conflict with the observed FIR EBL peak. Third, to obtain consistency with the CCSN rates, one would need to either increase the mass range for CCSN or invoke almost negligible true dust correction, and neither seems plausible.

Therefore, we conclude that the CSFH cannot be smaller by even a factor $\sim$ 2; the maximal reduction factor is perhaps $78/50 \sim 1.5$. This sets the lower limit of the CSFH normalization. We remind the reader that the DSNB detectability is not directly affected by the value of this factor, since the flux is directly normalized by the CCSN rate and is independent of uncertainties associated with low-mass star formation. 

\section{DSNB Predictions \label{sec:dsnb}}
In the previous section we explored the consistent picture of stellar birth, life, and death. As a result, we obtained the fiducial astrophysical input and uncertainties. In this section, we introduce the DSNB and discuss the neutrino emission per supernova and neutrino properties, i.e., the emission inputs and uncertainties. We then make predictions for the DSNB flux. 

\subsection{DSNB formalism}
The predicted DSNB number flux, over $4\pi$, is calculated by integrating $R_{\rm CCSN}(z)$ multiplied by the neutrino emission per supernova, $\mathrm{d}N/\mathrm{d}E$, appropriately redshifted, over cosmic time \cite{Ando:2004hc},
\begin{equation} \label{DSNB}
\frac{d\phi(E)}{dE} = c \int R_{\rm CCSN}(z)
\frac{dN(E^\prime)}{dE^\prime} (1+z) \left| \frac{dt}{dz} \right| dz,
\end{equation}
where $E^\prime = E (1+z)$. 

Progenitors over a wide range of masses lead to similar neutron star masses and hence neutrino emissions \cite{Takahashi:2003rn}. The dominant neutrino emission occurs during the Kelvin-Helmholtz cooling phase, when the newly formed hot and dense protoneutron star cools to a neutron star \cite{Burrows:1986me} (see also Refs.~\cite{Raffelt:book} and \cite{Kotake:2005zn}). Neutrinos and anti-neutrinos of all flavors are produced ($\nu_e$, $\bar{\nu}_e$, and $\nu_x$; where $\nu_x$ refers to $\nu_\mu$, $\nu_\tau$, and their antiparticles), and each species carries away an approximately equal fraction of the total energy, $E_\nu^{\mathrm{tot}} \approx 3\times 10^{53}$ erg. Their spectra are to a good approximation thermal; we summarize some temperatures from numerical supernova simulations in Table \ref{table:temperature} (where we use $i$ to explicitly denoted quantities at production). The hierarchy $T_{\nu_e}^i<T_{\bar{\nu}_e}^i<T_{\nu_x}^i$ reflects the different radii at which each neutrino species decouples, which in turn arises from the relevant neutrino interactions. 

The observed $\bar{\nu}_e$ spectra outside the star are linear combinations of the neutrino spectra at production, owing to neutrino mixing. The time-integrated $\bar{\nu}_e$ spectrum per supernova is well approximated by the Fermi-Dirac distribution with zero chemical potential \cite{Raffelt:book,Kotake:2005zn},
\begin{equation} \label{spectrum}
\frac{dN_{\bar{\nu}_e}}{dE^\prime_{\bar{\nu}_e}}(E^\prime_{\bar{\nu}_e})=
\frac{E_\nu^\mathrm{tot}}{6} 
\frac{120}{7\pi^4} 
\frac{E^{^\prime 2}_{\bar{\nu}_e}}{T^4_{\bar{\nu}_e}}
\left( e^{E^\prime_{\bar{\nu}_e}/T_{\bar{\nu}_e}}+1 \right)^{-1},
\end{equation}
where $T_{\bar{\nu}_e}$ is the effective $\bar{\nu}_e$ temperature outside the star after neutrino mixing. This is the temperature that is measured by neutrino detectors and we therefore use it for predictions. The effective temperature contains information on stellar and neutrino physics, and it is a separate problem to work backwards from the effective spectrum to the initial spectra, taking into account the effects of neutrino mixing. 

\subsection{Neutrino emission per supernova}\label{neutrinoemission}
It is important to address the range of expected neutrino emission per supernova, because we integrate over the entire CCSN population. Potential processes that affect the neutrino emission are both microphysical \cite{Pons:1998mm,Horowitz:2004yf,Burrows:2004vq,Langanke:2007ua} and macrophysical \cite{Thompson:2004if,Beacom:2000qy,Sumiyoshi:2007aa,Sumiyoshi:2008zw}. Neutrino mixing also plays an important role \cite{Takahashi:2001ep,Takahashi:2002cm,Ando:2002zj}. Although a body of predictions is given in the literature, it should be emphasized that numerical simulations usually do not explode, and that even the most state-of-the-art simulations do not reach beyond the accretion phase at a few hundred milliseconds after bounce. We therefore try to be as general as possible and focus on the time-integrated emission needed for the DSNB.

\begin{table} 
\caption{Flavor-dependent temperatures from some examples of numerical supernova simulations in the literature.}\label{table:temperature}
\begin{ruledtabular}
\begin{tabular}{lccccc}
Author & time\footnotemark[1] & $T_{\nu_e}^i$\footnotemark[2] & $T_{\bar{\nu}_e}^i$\footnotemark[2] & $T_{\nu_x}^i$\footnotemark[2] & Ref. \\
\hline 
Myra \& Burrows (1990)      & 0.2  & 3.3 & 4.0 & 8.0 & \cite{Myra:1990tt} \\
Totani et al.~(1998)        & 0.5  & 3.9 & 4.9 & 6.3 & \cite{Totani:1997vj} \\
                            & 10   & 3.5 & 6.3 & 7.9 & \\
Rampp \& Janka (2000)       & 0.5  & 2.3 & 3.4 &     & \cite{Rampp:2000ws} \\
Liebendoerfer et al.~(2001) & 0.5  & 4.2 & 4.6 & 5.3 & \cite{Liebendoerfer:2000cq} \\
Mezzacappa et al.~(2001)    & 0.5  & 3.6 & 4.2 & 5.2 & \cite{Mezzacappa:2000jb} \\
Keil et al.~(2003)          &      & 3.7 & 4.0 & 5.2 & \cite{Keil:2002in} \\
Thompson et al.~(2003)      & 0.2  & 2.9 & 3.5 & 4.5 & \cite{Thompson:2002mw} \\
Liebendoerfer et al.~(2004) & 0.5  & 3.7 & 4.2 & 4.5 & \cite{Liebendoerfer:2002xn} \\
                            & 0.5\footnotemark[3]  & 4.5 & 4.8 & 7.5 &  \\
Sumiyoshi et al.~(2007)     & 1.4\footnotemark[3]  & 6.6 & 7.0 & 10  & \cite{Sumiyoshi:2007aa} \\
\end{tabular}
\end{ruledtabular}
\footnotetext[1]{postbounce time in seconds}
\footnotetext[2]{temperatures at production in MeV}
\footnotetext[3]{leads to black hole formation}
\end{table} 

\subsubsection{Total neutrino energies}
The total energy budget in all flavors of neutrinos is dictated by the binding energy of the final remnant,
\begin{equation} \label{binding}
E_\nu^{\mathrm{tot}} \simeq E_{\rm bind} = 3 \times 10^{53} 
\left( \frac{M_{NS}}{1.4 \, \mathrm{M_\odot}} \right)^2 
\left( \frac{R_{NS}}{10\,\mathrm{km}} \right)^{-1} 
\mathrm{erg}, 
\end{equation}
where $M_{NS}$ and $R_{NS}$ are the neutron star mass and radius. The neutron star mass is best estimated by measurements in high-mass X-ray binaries and binary pulsars containing a radio pulsar and a neutron star companion. The most likely value is $\approx$ 1.4 $\mathrm{M_\odot}$, with a range 1.2--1.6 $\mathrm{M_\odot}$ \cite{Lattimer:2006xb}. The radii are more difficult to measure, but are consistent with being $\sim$ 10 km. Note the higher masses measured in low-mass X-ray binaries are attributed to their longer mass accretion histories.

The $\bar{\nu}_e$ emission depends on how $E_\nu^{\mathrm{tot}}$ is partitioned between neutrino flavors. In the numerical supernova simulations of the Lawrence Livermore (LL) group, which successfully followed the simulation into the Kelvin-Helmholtz cooling phase ($\sim$ 18 s) \cite{Totani:1997vj}, almost exact luminosity equipartition is seen throughout. The time-integrated total $\bar{\nu}_e$ energy is $4.7 \times 10^{52}$ erg, in good agreement with $E_\nu^{\mathrm{tot}}/6$. 

While the LL group succeeded in obtaining an explosion, it has been realized that they lacked neutrino processes now recognized as important. A review of the literature shows that luminosity equipartition is not universal: while $L_{\nu_e}^i \approx L_{\bar{\nu}_e}^i$ appears to be robust, $L_{\nu_x}^i$ varies by a factor 2--3 in either direction, depending on the evolutionary phase and on numerical methods. However, we remind the reader that these simulations only reach a few hundred milliseconds at most, when the bulk of neutrinos have not yet been emitted, and therefore say little about energy equipartition.

In addition, the observed $\bar{\nu}_e$ emission is a linear combination of $\bar{\nu}_e$ and $\bar{\nu}_x$ at production, due to neutrino mixing. Mixing scenarios, which are dependent on the neutrino mass hierarchy and oscillation parameters as well as the neutrino and stellar densities through which the neutrinos propagate, have been studied systematically \cite{Takahashi:2001ep,Takahashi:2002cm,Ando:2002zj,Dasgupta:2007ws,EstebanPretel:2007yq,Chakraborty:2008zp}. These give the effective $\bar{\nu}_e$ temperature a value in between the extremes corresponding to a pure $\bar{\nu}_e$ at production and a pure $\bar{\nu}_x$ at production.

Observationally, analysis of $\bar{\nu}_e$'s from SN 1987A show that the observed $\bar{\nu}_e$ energy budget is 3--$6 \times 10^{52}$ erg \cite{Arafune:1987ua,Janka:1989aa}, confirming the approximate energy equipartition after neutrino mixing. Variations away from energy equipartition at production are likely reduced by the effects of neutrino mixing as described above.

In some cases, the total energy budget may be smaller than Eq.~(\ref{binding}). One example is the effect of the equation of state for dense matter. Another possibility would seem to be the case of prompt black hole formation, where the neutrino signal is expected to be abruptly cut short \cite{Baumgarte:1996iu,Liebendoerfer:2002xn}. However, recent simulations show that black hole formation is preceded by an increase in the neutrino luminosity \cite{Sumiyoshi:2007aa,Sumiyoshi:2008zw,Fischer:2008rh}, as the gravitational energy of the rapidly accreting matter is released. The time integrated neutrino luminosity can be as high or higher than in normal neutron star formation. Another potential case is rotation, which could decrease Eq.~(\ref{binding}) by bloating the neutron star, although the effects are most likely not significantly large for an integrated population (see the light curves converging with time in Fig.~2 of Ref.~\cite{Thompson:2004if}). These are all interesting physics to study using supernova neutrinos.

\subsubsection{Average neutrino energies}
In general, the neutrino temperature is only weakly dependent on the neutron star parameters. Assuming a thermal spectrum, the neutrino luminosity is $L_\nu \propto R_\nu^2 T_\nu^4$. Since the total energy budget is well defined, it is useful to consider $E_\nu^{\mathrm{tot}} \approx L_\nu \Delta t$, where $\Delta t \approx 10 \, \mathrm{s}\, (R_{NS}/10\,\mathrm{km})^2 (\rho/2 \rho_{nm})^{2/3}$ is the time scale for the neutrino cooling phase \cite{Burrows:1984aa,Burrows:1990ts} and $\rho_{nm}$ is the nuclear density. This gives $T_\nu \sim 5 \, (M_{NS}/1.4 \, \mathrm{M_\odot})^{1/3} (R_{NS}/10\,\mathrm{km})^{-3/4}$ MeV. Although this is a simplistic analysis, it shows that $T_\nu$ is weakly dependent on the neutron star mass and radius. For example, the neutrino emission from a more massive neutron star-neutron star merger system is comparable to that of a normal supernova \cite{Rosswog:2003rv}.

Indirect constraints on the $\nu_x^i$ spectrum have been placed from the chemical abundance of $\nu_x$-process induced elements. Since the inelastic $\nu_x$ scattering off nuclei uniquely leads to the production of specific isotopes \cite{Heger:2003mm}, the required $\nu_x$ flux can be inferred. For $E_{\nu_x}^i=4$--$5.8 \times 10^{52}$ erg, $T^i_{\nu_x}$ is constrained to lie between 4--7 MeV \cite{Yoshida:2005uy,Yoshida:2008zb}.

The DSNB $\bar{\nu}_e$ spectrum can be calculated from the observed SN 1987A $\bar{\nu}_e$ \cite{Fukugita:2002qw,Lunardini:2005jf,Yuksel:2007mn}. The SN 1987A spectrum constructed by fitting \cite{Lunardini:2005jf} or by nonparametric methods \cite{Yuksel:2007mn} shows similar distortions from a thermal spectrum. For the purposes of DSNB detection, the spectrum near detector threshold is important, which as we discuss below is $\approx 20$ MeV for the current SK and $\approx 10$ MeV in a gadolinium-enhanced SK. The spectrum of Ref.~\cite{Yuksel:2007mn} is similar to a 5 MeV thermal spectrum for high energies ($E_\nu \gtrsim 30$ MeV), and closer to a 4 MeV thermal spectrum at somewhat smaller energies ($E_\nu \lesssim 30$ MeV). Therefore, the SN 1987A spectrum could be treated as similar to a 4 MeV thermal spectrum, which is on the low end of the theory predictions. The limited sample of the SN 1987A data means this spectrum serves as a guidance rather than the definitive neutrino spectrum per supernova; there is no good reason to invoke a smaller spectrum.

\begin{figure}[t]
\includegraphics[width=3.25in]{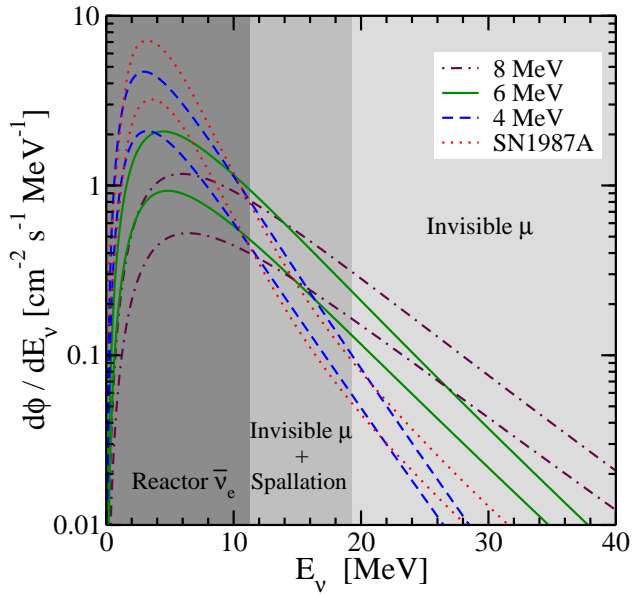}
\caption{\label{fig:DSNBflux} DSNB flux spectrum for emitted neutrino spectra as labeled. For each spectrum, two curves are plotted representing the full range of uncertainties due to astrophysical inputs (the fiducial prediction lies in between). The shadings indicate backgrounds, with origins as labeled. Decays of invisible muons and spallation products would be reduced in a gadolinium-enhanced SK, opening the energy region 10 MeV and above to a rate-limited DSNB search; see Fig.~\ref{fig:DSNBevent}.}
\end{figure}

\subsection{DSNB flux prediction} \label{DSNBflux}
In Fig.~\ref{fig:DSNBflux} we show the predicted DSNB flux for a selection of neutrino spectra as labeled. For each, the two curves correspond to the upper and lower $R_{\rm CCSN}$ inputs from Fig.~\ref{fig:CCSNrate}. The fiducial astrophysical input lies in between. The figure illustrates the relative sizes of the DSNB uncertainties originating from the astrophysics and emission inputs. The shading represents the relevant backgrounds which detections of the DSNB must compete against. We discuss in Sec.~\ref{sec:future} how these backgrounds would be rejected in a gadolinium-enhanced SK.

We adopt thermal spectra given by Eq.~(\ref{spectrum}) with effective temperatures of 4, 6, and 8 MeV. These are effective temperatures, reflecting the range of initial temperatures shown in Table \ref{table:temperature}, as well as the range of neutrino mixing effects described in section \ref{neutrinoemission}. We also show the neutrino spectrum directly reconstructed from the SN 1987A neutrino data \cite{Yuksel:2007mn}.


We see that with the current astrophysical uncertainties, the emission uncertainties dominate at $E_\nu \gtrsim 20$ MeV. The DSNB in this energy range therefore provides information on the emission inputs. At lower energy the fluxes and uncertainties converge. 

\section{DSNB constraints and detection \label{sec:sk}}

\subsection{Super-Kamiokande constraints}

\begin{figure}[t]
\includegraphics[width=3.25in]{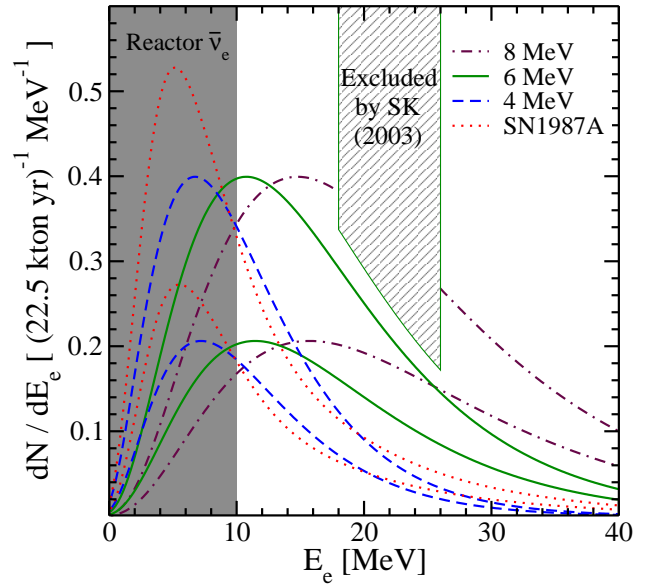}
\caption{\label{fig:DSNBevent} DSNB event rates at SK (flux spectra weighted with the detection cross section) against positron energy. Note the linear axis. We hatch in the 2003 upper limit by the Super-Kamiokande Collaboration, $< 2$ events (22.5 kton yr)$^{-1}$ in the energy range 18--26 MeV. The limit applies to all spectra (see text). In a gadolinium-enhanced SK, decays of invisible muon and spallation products would be reduced, opening up the energy range $\gtrsim$ 10 MeV for DSNB search (unshaded region).}
\end{figure}

Super-Kamiokande is a water \v{C}erenkov neutrino detector with a fiducial volume of 22.5 kton, sensitive to DSNB through the inverse beta reaction $\bar{\nu}_e p \to e^+ n$ on free protons (hydrogen nuclei). The positron energy faithfully represents the neutrino energy, and $E_e \simeq E_{\bar{\nu}_e}-1.3 \,\mathrm{MeV}$ to zeroth order in $M_p$. In 2003, the Super-Kamiokande Collaboration reported a stringent upper limit on the integrated DSNB $\bar{\nu}_e$ signal above $E_e=18$ MeV, using 1496 days (4.1 yr) of data. The dominant background in this energy range was the decay of invisible muons, the fixed spectrum of which rises steeply with positron energy \cite{Kaplinghat:1999xi,Malek:2002ns}. The best signal to noise ratio is therefore obtained at the lowest energy bin, 18--26 MeV, where the background rate is some $\sim 3$ events yr$^{-1}$. By searching for DSNB neutrinos over detector backgrounds, the Super-Kamiokande Collaboration has limited the DSNB flux to be $\phi(E_{\bar{\nu}_e} > 19.3 \,\mathrm{MeV}) < 1.2$ cm$^{-2}$ s$^{-1}$. In a more physically relevant form, the limit is $\lesssim$ 2 events (22.5 kton yr)$^{-1}$ in the lowest energy bins \cite{Malek:2002ns,Yuksel:2005ae}. 

\subsection{DSNB event spectrum}
In Fig.~\ref{fig:DSNBevent} we show the predicted event rates (flux spectrum weighted with the detection cross section \cite{Vogel:1999zy,Strumia:2003zx}) as a function of positron energy, which is the observed quantity in SK. We account for corrections of order $1/M_p$ in both the cross section and kinematics, which results in a net $\sim$ 20\% reduction in event rates at the current SK (and 10\% reduction at a gadolinium-enhanced SK). While the SK limit could be weakened by these corrections \cite{Lunardini:2008xd}, we adopt the published limit \cite{Malek:2002ns}, since the net effect of the corrections is smaller than the statistical uncertainties on the number of background events at SK. Furthermore, corrections should also be applied to the source, which would reduce the neutrino opacity and lead to more high-energy neutrino escaping, partially canceling the effect of the corrections. One should also consider that there are uncertainties in approximating the neutrino spectrum, which will not be purely thermal. 

We show four DSNB spectra in the figure, as labeled. We also show the region already excluded by SK \cite{Malek:2002ns,Yuksel:2005ae} by hatched shading. The exclusion region is drawn by scaling the 6 MeV spectra up so that the predicted event rate in the energy range 18--26 MeV is 2 events (22.5 kton yr)$^{-1}$. Since SK currently probes a narrow range of the exponential tail of the thermal neutrino spectra, the exclusion region has an almost temperature-independent shape. 

The SK limit already partially excludes the dominance of $\bar{\nu}_e$ effective temperatures at or above 8 MeV. The 6 MeV DSNB, which is typical of scenarios with neutrino mixing, is within at most a factor 2 of the current SK limit. Furthermore, the 4 MeV spectrum, which reflects our lower estimate for the effective temperature, lies within a factor 4 of the current SK limit. Effective temperatures in the relevant energy range lower than 4 MeV are unlikely given the SN 1987A data and general considerations above. In Table \ref{table:counts} we show the predicted event numbers in the energy bin 18--26 MeV, which can be compared to the SK event limit, and which can be improved soon. 

\begin{table} 
\caption{Integrated DSNB event rates in the positron energy range 18--26 MeV for the current SK and 10--26 MeV for a gadolinium-enhanced SK. The uncertainties reflect the upper and lower $R_{\rm CCSN}$, defined from the CSFH.}\label{table:counts}
\begin{ruledtabular}
\begin{tabular}{ccc}
$\bar{\nu}_e$ spectrum & 
   \multicolumn{2}{c}{events [(22.5 kton yr)$^{-1}$]} \\
 & $10<E_e/\mathrm{MeV}<26$ & $18<E_e/\mathrm{MeV}<26$ \\
\hline 
8 MeV     & $4.2 \pm 1.4$      & $2.0 \pm 0.7$      \\
6 MeV     & $3.5 \pm 1.1$      & $1.3 \pm 0.4$      \\
4 MeV     & $1.8 \pm 0.5$      & $0.4 \pm 0.1$      \\
SN 1987A  & $1.7 \pm 0.5$      & $0.5 \pm 0.1$      \\
\end{tabular}
\end{ruledtabular}
\end{table} 

\subsection{DSNB future prospects} \label{sec:future}
In the current SK, DSNB $\bar{\nu}_e$ are detected in a singles positron search, $\bar{\nu}_e p \to e^+ n$, for which there are very large background rates \cite{Malek:2002ns}. The largest background above $\sim$ 18 MeV comes from the decay of invisible muons, i.e., non-relativistic muons produced from atmospheric $\nu_\mu$ and $\bar{\nu}_\mu$ \cite{Kaplinghat:1999xi}. The electrons and positrons produced in muon decays cannot be distinguished from those of the signal inverse beta decay reaction. At energies below $\sim$ 18 MeV, the decays of spallation products of cosmic-ray muons also contribute to the background (see Fig.~\ref{fig:DSNBflux}).

With 0.2\% dissolved gadolinium, the neutrons produced in inverse beta decay can be identified with high efficiency, via the photons emitted upon neutron capture on gadolinium. This delayed signal allows tight temporal and spatial coincidence for signal events, reducing the invisible muon background by a factor $\sim$ 5 and removing the spallation backgrounds in the range 10--18 MeV, opening it up to DSNB searches \cite{Beacom:2003nk,Yuksel:2005ae}. This is shown in Fig.~\ref{fig:DSNBevent} where the relevant shadings shown in Fig.~\ref{fig:DSNBflux} have been removed. Below 10 MeV, reactor $\bar{\nu}_e$ still overwhelm the DSNB signal \cite{Beacom:2003nk}. 

Importantly, the enhancement allows a mostly rate-limited, rather than background-limited, DSNB search, so that the sensitivity improves linearly with exposure. Another important point is that a gadolinium enhancement could be applied to the existing SK detector \cite{Beacom:2003nk}.

The advantages of an enhanced SK are dramatic. At the lowered energy threshold, the event spectrum and uncertainties converge, so that the total range in predictions is only a factor $\sim$ 2 (i.e., from the lowest predicted 4 MeV curve to the highest predicted 6 MeV curve). The predicted total event rate in the energy range 10--26 MeV is almost triple that in SK with a high energy threshold, since the peak of the DSNB spectra can be probed; a gadolinium-enhanced SK is almost guaranteed to detect the DSNB. A non-detection would require novel stellar or neutrino physics, for example a change in the equation of state, extremely fast rotation, invisible neutrino decays on cosmological scales \cite{Fogli:2004gy,Ando:2003ie}, or the effect of hypothetical particles on the emission model of supernovae. Any such explanations would have to show why the SN 1987A detection results were atypical. 

\section{Discussions and Conclusions \label{sec:conclusion}}
Neutrinos are the only probes (possibly apart from gravitational waves) of the central regions of core-collapse events, and their study and detection are strongly motivated by many areas of physics. In this paper we assess the uncertainties for DSNB searches and implications for stellar and neutrino physics. The results can largely be divided into three categories.

\subsection{On astrophysical inputs}
We start with an up-to-date compilation of the CSFH and cross-check it with other observables. While our aim is to constrain the astrophysical inputs for the DSNB (i.e., the CCSN rate), these new results are of value in their own right.

\begin{itemize}
\item \emph{Consistency with CCSN rates}: Using a collection of recently observed CCSN rates, we show that they are in agreement with predicted CCSN rates from the CSFH, up to $z \sim 1$. Importantly, this check is almost independent of the IMF.
\item \emph{Consistency with EBL}: Using new EBL measurements, including TeV gamma-ray constraints, we show that predictions from the CSFH are in agreement with observations. The observed total EBL is $73^{+21}_{-26} \, \mathrm{nW \, m^{-2} \, sr^{-1}}$, while we predict $78^{+31}_{-24} \, \mathrm{nW \, m^{-2} \, sr^{-1}}$ for the BG IMF.
\item \emph{Consistency with stellar mass density}: The choice of the BG IMF is independently supported by studies of the stellar mass density, which show agreement in $z \lesssim 0.7$.
\end{itemize}

In conclusion, our fiducial CSFH together with the BG IMF gives excellent agreement among all observations considered, preparing the astrophysical inputs for the DSNB. In particular, we stress how the astrophysical inputs cannot be too low, in order to maintain the self-consistent picture of the birth, life, and death of stars, as illustrated by the cross-checks.

\subsection{On emission inputs}
Next we assess the range of neutrino emission expected from supernovae. Coupled to our cross-checked astrophysical inputs, this allows us to discuss how constraining the present limit on the DSNB is. Our best-determined results are shown in Fig.~\ref{fig:DSNBevent}.

\begin{itemize}
\item \emph{Neutrino emission}: Ultimately, this is a quantity to be studied using supernova neutrino detections. We thus discuss the generic emission of $\bar{\nu}_e$ from supernovae using simple arguments based on the energetics of core collapse, neutrino mixing, and the observed SN 1987A neutrino burst. We also show the neutrino temperature depends weakly on the remnant parameters. We conclude that the relevant high-energy $\bar{\nu}_e$ emission cannot be made too small. 
\item \emph{Present constraints}: The SK limit is already probing interesting parameter regions. The dominance of an effective $\bar{\nu}_e$ temperature above 8 MeV is constrained. Similarly, a high rate of dark core collapses is also prohibited by an amount that depends on the assumed temperature.
\item \emph{Implications for stellar and neutrino physics}: If an effective temperature is ruled out, this rules out neutrino mixing and initial neutrino temperatures that would lead to that effective temperature in the DSNB energy range. The correspondance between the initial and effective neutrino emissions is dependent on the neutrino mixing scenario, which has been studied systematically by various authors.
\end{itemize}

To conclude, from general and SN 1987A considerations, the neutrino emission cannot be too low. The current SK limit is already constraining interesting neutrino effective temperatures. Noting the SK limit was placed in 2003, more data and improved cuts will allow better sensitivity.

\subsection{On detection in SK}
With the DSNB inputs and their uncertainties checked, we discuss implications for DSNB detection in the future. 

\begin{itemize}
\item \emph{SK prospects}: The DSNB is near detection in SK. The 6 MeV spectrum, typical of scenarios with neutrino mixing, is within a factor $\sim$ 2 of the current SK limit. The factor increases to $\sim$ 4 for the lower 4 MeV and SN 1987A reconstructed spectra.
\item \emph{SK with gadolinium}: Intriguingly, the fluxes and uncertainties converge at the improved detection threshold (\mbox{$\sim$ 10 MeV}), so that predictions span an uncertainty of a factor $\sim$ 2. The predicted event rate between 10--26 MeV is 1.2--5.6 events yr$^{-1}$, leaving no room for the DSNB to escape detection.
\item \emph{Future physics with SK}: The effective temperature $T_{\bar{\nu}_e}$ contains physics concerning neutrino emission from the collapsed core and on neutrino mixing. Stellar and neutrino physics can be extracted by future comparisons between the measured and theoretical $T_{\bar{\nu}_e}$. Future SK analysises should report results directly in terms of the time-integrated luminosity and effective temperature \cite{Yuksel:2005ae}, using the astrophysically-measured supernova rate.
\end{itemize}

To conclude, while the current SK will continue probing interesting physics, a gadolinium-enhanced SK is almost guaranteed to detect DSNB events. A non-detection would require novel stellar or neutrino physics. Together with the decreasing astrophysical uncertainties, these results strongly support the imminent detection of the DSNB and solidify its important role in understanding supernova and neutrino physics.

\section*{Acknowledgments}

We thank Shin'ichiro Ando, Maria Terese Botticella, Thomas Dahlen, Andrew Hopkins, Cecilia Lunardini, Katsuhiko Sato, Stephen Smartt, Todd Thompson, Stephen Wilkins, and Mark Vagins for helpful discussions; Matt Kistler and Hasan Yuksel for helpful discussions and technical assistance. S.H.~thanks the hospitality of CCAPP, Ohio State University, where this work took place. S.H.~was supported by CCAPP, J.F.B.~was supported by NSF CAREER Grant PHY-0547102, and E.D.~was partially supported by NASA grant LTSA 03-0000-065.


\bibliographystyle{h-physrev4}
\bibliography{bibliography}


\end{document}